\documentclass{pasj00}
\draft

\newcommand{\eq}{equation}

\makeatletter
\def\lsim{\mathrel{\mathpalette\gl@align<}}
\def\gsim{\mathrel{\mathpalette\gl@align>}}
\def\gl@align#1#2{\lower.6ex\vbox{\baselineskip\z@skip\lineskip\z@
    \ialign{$\m@th#1\hfil##\hfil$\crcr#2\crcr\sim\crcr}}}
\makeatother

\begin{document}
\SetRunningHead{Madokoro, Motizuki, \& Shimizu}{Importance of Prolate
Neutrino Radiation in Core-Collapse Supernovae: The Reason for the Prolate
Geometry of SN1987A?}
\Received{2003/12/26}
\Accepted{2004/5/25}

\title{Importance of Prolate Neutrino Radiation in \\
Core-Collapse Supernovae: The Reason for \\
the Prolate Geometry of SN1987A?}

\author{Hideki \textsc{Madokoro}}
\email{madokoro@postman.riken.jp}

\author{Tetsuya \textsc{Shimizu}}
\email{tss@postman.riken.jp}
\and
\author{Yuko {\sc Motizuki}}
\email{motizuki@riken.jp}

\altaffiltext{}{RIKEN, Hirosawa 2-1, Wako 351-0198, Japan}


\maketitle

\begin{abstract}
We have carried out 2-D simulations of core-collapse supernova
explosions.  The local neutrino radiation field is assumed to have its
maximum value either at the symmetry (polar) axis or on the equatorial
plane.  These lead to the prolate and oblate explosions, respectively.
We find that the gain of the explosion energy in the prolate explosion
evolves more predominately than that in the oblate one when the total
neutrino luminosity is given.  Namely, the prolate explosion is more
energetic than the oblate one.

One of the authors (Shimizu et al. 2001) showed for the first time that
globally anisotropic neutrino radiation produces more powerful explosion
than the spherical neutrino radiation does.  In our previous study
\citep{MaShMo03}, we improved the numerical code of Shimizu et al. and
demonstrated that the globally anisotropic neutrino radiation yields
more energetic explosion than spatially-fluctuated neutrino radiation
does.  Together with the result of this paper, we conclude that the
globally anisotropic (prolate) neutrino radiation is the most effective
way of increasing the explosion energy among various types of explosions
investigated in these studies.  We discuss the reason for this.  Our
result is suggestive of the fact that the expanding materials of SN1987A
is observed to have a prolate geometry.
\end{abstract}

\section{Introduction}
Aspherical explosion is one of the key issues in the study of
core-collapse supernovae.  Observations have been suggested that the
spherical symmetry is broken in several supernova explosions.  For
example, speckle and spectropolarimetry observations have shown that
the spectra of core-collapse supernovae are significantly polarized at
a level of 0.5 to 4\% \citep{Me88,Je91,Wa96,Le00,Wa01,Ho02}.  Recent
Hubble Space Telescope images and spectroscopy have revealed that
SN1987A has an axially symmetric geometry \citep{Wa02}.  \citet{Wa02}
also reported that the polarization represents a prolate geometry which
has been fixed in time.  They discussed that a jet-driven model based
on asymmetry associated with neutrino flow
\citep{ShYaSa94,FrHe00,ShEbSaYa01} may account for the observed
asymmetries of SN1987A.  Large values of observed pulsar kicks ( $>$
400 km\ s$^{-1}$; see, e.g., Fryer, Burrows, \& Benz 1998) are also a
manifestation of asymmetry.  Furthermore, it has been theoretically
demonstrated that most simulations assuming spherical symmetry fail to
yield robust explosions (e.g., Liebend\"orfer et al. 2001).  These
facts necessarily lead us to multi-dimensional simulations.

Several groups have performed 2-D and 3-D simulations.  It was shown
\citep{MiWiMa93,HeBeHiFrCo94,BuHaFr95,JaMue96,Me98,FrHe00,FrWa02,
KiPlJaMu03} that convection in the neutrino-heated region behind the
supernova shock or inside a nascent neutron star plays an important
role to increase the explosion energy.

Some groups also noticed the role of anisotropic neutrino emission from
a protoneutron star.  Janka and M\"onchmeyer (1989a,b) carried out 2-D
simulations with rotation in order to explain the observed properties
of SN1987A.  When a protoneutron star rotates, centrifugal force will
deform the core into an oblate form, and hence the neutrino flux
distribution will preferentially be concentrated on the rotational
(polar) axis.  They concluded that the neutrino flux along the
rotational axis could become up to three times larger than that on the
equatorial plane.  Later, \citet{KeJaMue96} performed 2-D simulations
without rotation but taking into account the convection inside the
protoneutron star (Ledoux convection).  They found that the neutrino
flux on the surface of a protoneutron star shows a fluctuated
anisotropy of $\sim$ 3-4 \%.  Recently, \citet{KoYaSa03} performed
simulations of the rotational core-collapse and obtained anisotropic
neutrino heating, depending on their initial model with the degree of
anisotropy in the neutrino heating rate from close to unity to more than
a factor of 10.

Shimizu et al. (1994, 2001) performed a series of systematic 2-D
simulations in order to study the effects of anisotropic neutrino
radiation on the explosion mechanism itself.  They found that, when
the total neutrino luminosity is given, only a few percent of enhancement
in the neutrino flux along the axis of symmetry is sufficient to increase
the explosion energy by a large factor.  They also found that this effect
saturates around a certain degree of anisotropy.  Furthermore, it was
found that the effect of anisotropic neutrino radiation becomes more
prominent when the total neutrino luminosity is low, while the
difference between the anisotropic and spherical models is significantly
diminished if the total neutrino luminosity is sufficiently high.
Note that they proposed a rotation as one of the
origins of anisotropic neutrino radiation; anisotropy in neutrino
radiation may also be originated from convection inside the nascent
neutron star, or asymmetric mass accretion onto the protoneutron star.

\citet{ShEbSaYa01} considered only the case in which the maximum value
in the neutrino flux distribution was located at the axis of
symmetry.  \citet{JaKe97}, however, pointed out that the neutrino flux
can also be peaked on the equatorial plane.  Following their previous
study (Keil et al. 1996), they carried out 2-D simulations with both
rotation and convection above and inside a nascent neutron star.  It
was found, at a very early phase of neutrino emission, that the
neutrino flux was peaked at the rotational axis.  However, they found
that at a later stage of evolution Ledoux convection occurs strongly on
the equatorial plane while suppressed near the rotational axis due to a
combination effect of centrifugal force and convection.  Consequently,
the neutrino flux on the equatorial plane was enhanced over that along
the rotational axis.

In this paper, we are going to study the effects of the oblate neutrino
heating, suggested by \citet{JaKe97}, on the explosion dynamics.  We
focus the shock position, the explosion energy, and the asymmetric
explosion.  Our purpose in this article, which follows the previous
studies \citep{ShEbSaYa01,MaShMo03}, is to investigate what kind of
neutrino radiation is favorable for a successful explosion through a
parameter study when the total neutrino luminosity is given.

This paper is organized as follows.  In \S~2, we describe our numerical
simulation.  We then discuss our numerical results in \S~3.  The results
of the oblate explosion are compared with those of the prolate one.  We
shall find that the locally intense neutrino heating along the pole, which
leads to a prolate explosion, is more effective way of increasing the
explosion energy than the oblate heating when the total neutrino
luminosity is given.   Finally, a brief summary is presented in \S~4.

\section{Numerical Simulation}
We perform 2-D simulations by solving hydrodynamical equations in the
spherical coordinate ($r,\theta$).  A generalized Roe's method is used
to solve the general equations of motion.  Our computational region
ranges from 50 km to 10000 km in radius from the center of the
protoneutron star.  We start our calculations when a stalled shock
wave is formed
at $\sim$ 200 km.  The temperature on the neutrinosphere, $T_{\nu}$, and
the total neutrino luminosity, $L_{\nu}$, are assumed to be constant
in our simulations.  The details of our numerical technique, the EOS
and the initial condition, are described in \citet{ShEbSaYa01}.

We obtain the initially stalled shock by solving stationary hydrodynamic
equations with assuming spherical symmetry.  For the purpose of this
paper, this assumption is enough.  Note that in a simulation model by
other group (Kotake et al. 2003), the stalled shock wave can become
almost spherical even if the protoneutron star rotates with the
rotational velocity which the authors call `moderate' (and with weak
differential rotation).  We stop our calculations with constant $L_{\nu}$
at $t=500$ ms after the shock stagnation.  This is because recent one-
and two-dimensional simulations have shown that the timescale of decay
of the neutrino luminosity after the shock stall is about 500 ms
\citep{Bu93,Wo94,JaMue95,FrHe00}.

We fix the radius of the protoneutron star, $R_{\rm NS}$, to be 50 km
in our simulation.  The neutrinosphere actually shrinks as a result of
neutrino emission (Burrows et al. 1995).  However, the effect
of shrinking of the neutrinosphere is not important in our present study.
The reasons are as follows: Firstly, the degree of anisotropy in the neutrino
radiation at a distant point from the neutrinosphere (which we denote
`$c_{2}$') depends strongly on the degree of anisotropy on
the neutrinosphere (denoted by `$a$'), while is insensitive to $R_{\rm NS}$
(see equation (\ref{eqn:c2a}) later).  Secondly, the region in which we are
interested is 1000 km above the neutrinosphere.  We then expect that
small change of the radius of the neutrinosphere (from 50 km to 40 km; see
Burrows et al. 1995) does not affect the shock dynamics qualitatively during
the timescale of 500 ms on which we simulate.  Thirdly, we fix both the radius
of the protoneutron star and the neutrino energy (or equivalently the neutrino
temperature) in our simulation.  On the other hand, the neutrinosphere
actually shrinks as a result of neutrino emission, while at the same time
the neutrino energy increases.  It is therefore a good approximation
to consider that the rate of absorption of the neutrino energy on the shocked
matter is approximately constant during the short timescale of 500 ms.
We then think that the qualitative result is the same as our present result,
even if we include both the shrinking of the protoneutron star and
the increase of the neutrino energy.

We have improved (Madokoro et al. 2003) the numerical code of
\citet{ShEbSaYa01}: the cells in the $\theta$-direction were shifted
by half of the cell size \citep{Sh95} in order to avoid a numerical error
near the pole.  Note that this numerical error was not serious and minor
for the investigation of the explosion energy, but may affect the results
of nucleosynthesis.

The local neutrino flux is assumed as
\begin{\eq}
  l_\nu(r,\theta) = \frac{7}{16}\sigma T_{\nu}^{4} c_{1}
  \left(1+c_{2}\cos^{2}\theta\right)\frac{1}{r^{2}},
  \label{eqn:nuflux}
\end{\eq}

\noindent
where $\sigma$ is the Stefan-Boltzmann constant.
In equation (\ref{eqn:nuflux}),
$c_{2}$ is a parameter which is related to the degree of anisotropy in
the neutrino radiation.  In order to see the effect of anisotropic
neutrino radiation itself on the explosion, the value of $c_{1}$ is
calculated from given $c_{2}$ so as to adjust the total neutrino
luminosity to that in the spherical model at the same $T_{\nu}$.
The total neutrino luminosity is obtained by integrating equation
(\ref{eqn:nuflux}) over the whole solid angle,
\begin{\eq}
  L_{\nu} = \int r^{2}l_{\nu}(r,\theta) d\Omega
          = \frac{7}{16}\sigma T_{\nu}^{4}\;4\pi c_{1}
            \left(1+\frac{1}{3}c_{2}\right),
  \label{eqn:lnutotal}
\end{\eq}

\noindent
which is equated to that of spherical explosion with the same $T_{\nu}$,
\begin{\eq}
  L_{\nu}^{\rm sp} = \frac{7}{16}\sigma T_{\nu}^{4}\;4\pi
  R_{\rm NS}^{2}.
  \label{eqn:lnutotalsph}
\end{\eq}

\noindent
By comparing equations (\ref{eqn:lnutotal}) with (\ref{eqn:lnutotalsph}),
we obtain
\begin{\eq}
  c_{1}=\frac{R_{\rm NS}^{2}}{1+c_{2}/3}.
  \label{eqn:c1}
\end{\eq}

One can easily confirm that the neutrino flux along the pole ($l_{z}$)
and that on the equatorial plane ($l_{x}$) are proportional to
$c_{1}(1+c_{2})$ and $c_{1}$, respectively.  The degree of anisotropy
at a distant point far from the neutrinosphere, $l_{z}/l_{x}$, is then
given by
\begin{\eq}
  \frac{l_{z}}{l_{x}} = 1 + c_{2}.
  \label{eqn:anisotropy}
\end{\eq}

\noindent
Accordingly, the neutrino flux has its maximum at the pole or on the
equator when the sign of $c_{2}$ is positive or negative, respectively.

\begin{figure}
  \begin{center}
    \FigureFile(160mm,160mm){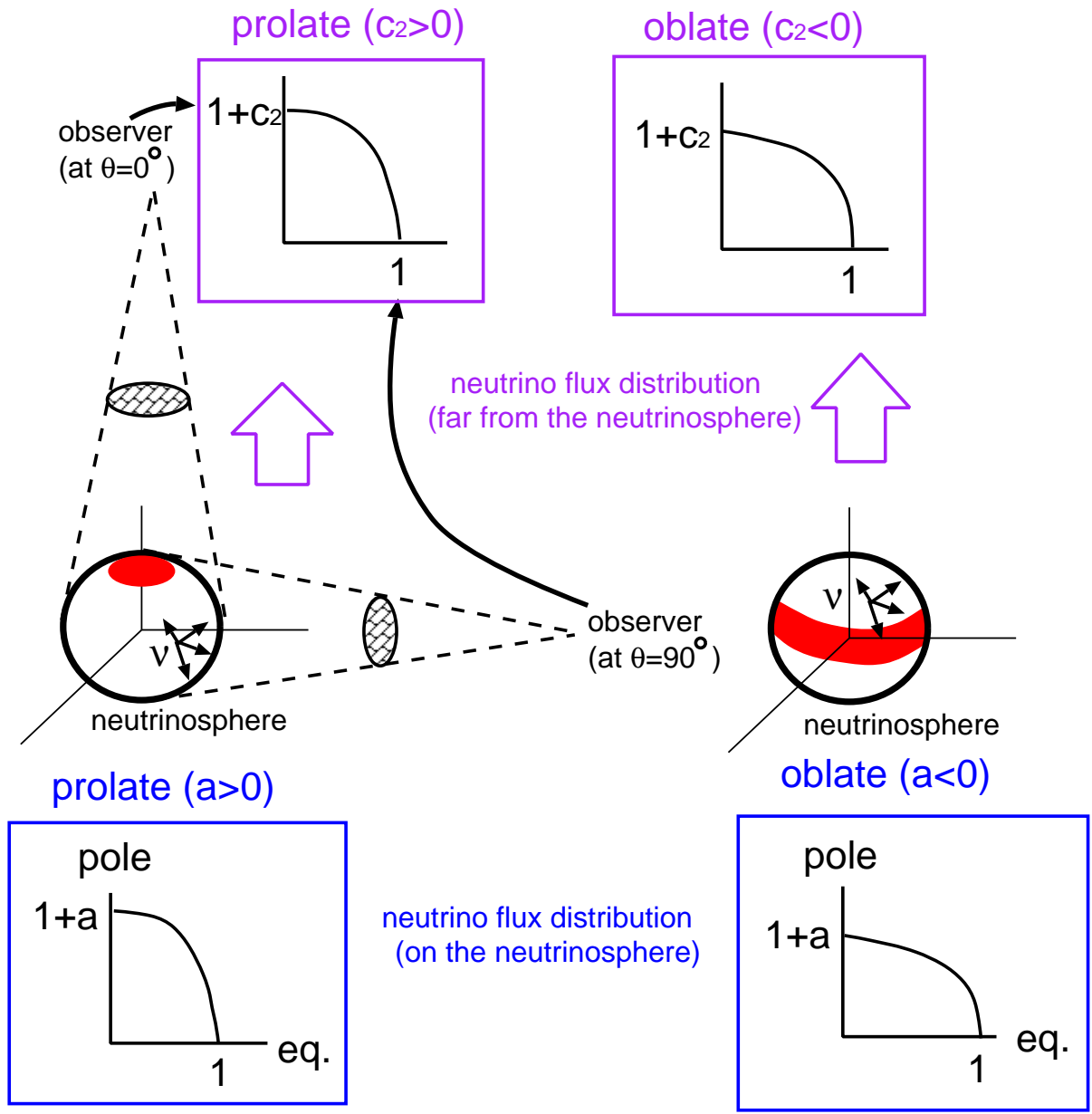}
  \end{center}
  \caption{A schematic picture which shows the relationship between the
  neutrino flux on the neutrinosphere and that for an observer far from
  the neutrinosphere.  Due to the geometric effect from an
  anisotropically radiating surface, $|c_{2}|$ is always smaller than
  $|a|$.}
  \label{fig:geo}
\end{figure}

We have to emphasize that the degree of anisotropy in the neutrino flux
distribution for an observer far from the neutrinosphere is different
from that on the neutrino-emitting surface.  This is schematically
illustrated in Fig.~\ref{fig:geo}.  When we observe the neutrino flux
far from the neutrinosphere, the local neutrino flux is seen as
equation (\ref{eqn:nuflux}).  On the other hand, the neutrino flux on
the neutrino-emitting surface has a similar profile but the degree of
anisotropy is different from equation (\ref{eqn:nuflux}).  This is
because a distant observer obtains the neutrino flux by integrating all
the contributions over the solid angle from an anisotropically
radiating surface, and hence the degree of anisotropy is reduced for
the observer (the geometric effect we call).  For the neutrino flux on
the neutrino-emitting surface, $c_{2}$ in equation (\ref{eqn:nuflux})
is replaced by $a$, where $a$ is a parameter which represents the
degree of anisotropy in the neutrino radiation on the neutrinosphere.
The angular dependence of the neutrino flux on the neutrinosphere is
then represented as
\begin{\eq}
  l_{\nu} \sim 1+a\cos^{2}\theta.
  \label{eqn:lnuonnsph}
\end{\eq}

In principle, the value of $c_{2}$ is calculated from a given $a$
taking into account the geometric effect from an anisotropically
radiating surface.
%
Although it is difficult to calculate the exact relationship between $c_{2}$
and $a$, we can estimate it in a same way as in Madokoro et al. (2003).
The neutrino flux observed far from the neutrinosphere is obtained by
averaging all contributions from the flux on the surface of the neutrinosphere.
From the similar procedure to that in Madokoro et al. (2003), we obtain
the ratio of the local
neutrino flux along the polar axis ($l_{z}$) to that on the equatorial plane
($l_{x}$) for an observer far from the neutrinosphere as a function of $a$,
\begin{\eq}
  \frac{l_{z}}{l_{x}}\sim \frac{4+2a}{4+a}.
  \label{eqn:lxlz2}
\end{\eq}

\noindent
By comparing equation (\ref{eqn:lxlz2}) with equation (\ref{eqn:anisotropy}),
we finally obtain
\begin{\eq}
  c_{2}\sim \frac{a}{4+a}.
  \label{eqn:c2a}
\end{\eq}

\noindent
Note that equation (\ref{eqn:lxlz2}) is different from equation (7) in
Madokoro et al. (2003).  This is because the profile of the neutrino flux
on the neutrinosphere is approximated by a step function in Madokoro et al.
(2003), while we directly use equation (\ref{eqn:lnuonnsph}) to obtain
(\ref{eqn:lxlz2}) in this paper.

\section{Prolate and Oblate Explosions}
We first study the prolate and oblate explosions with the same degree of
anisotropy; $l_{z}/l_{x}=1.1$ (model pro100) and $l_{z}/l_{x}=1/1.1$ (model
obl091).  The value of $c_{2}$ is 0.1 and $-0.091$, respectively.
The spherical model (model sph000) is also examined for comparison.
The temperature on the neutrino-emitting surface,
$T_{\nu}$, is at first fixed to be 4.70 MeV.  For models pro100, obl091,
and sph000, we also study the cases with different values of
the neutrino temperature ($T_{\nu}=4.65$ and $4.75$ MeV) in order to check
the sensitivity of the results on the neutrino temperature.  We further
consider several cases
changing the degree of anisotropy in the neutrino radiation in order to
examine at what degree of anisotropy the effect of anisotropic neutrino
radiation will saturates for both the prolate and oblate models.
The models we examined are summarized in Table~\ref{tab:first}.  

\begin{figure}
  \begin{center}
    \FigureFile(80mm,80mm){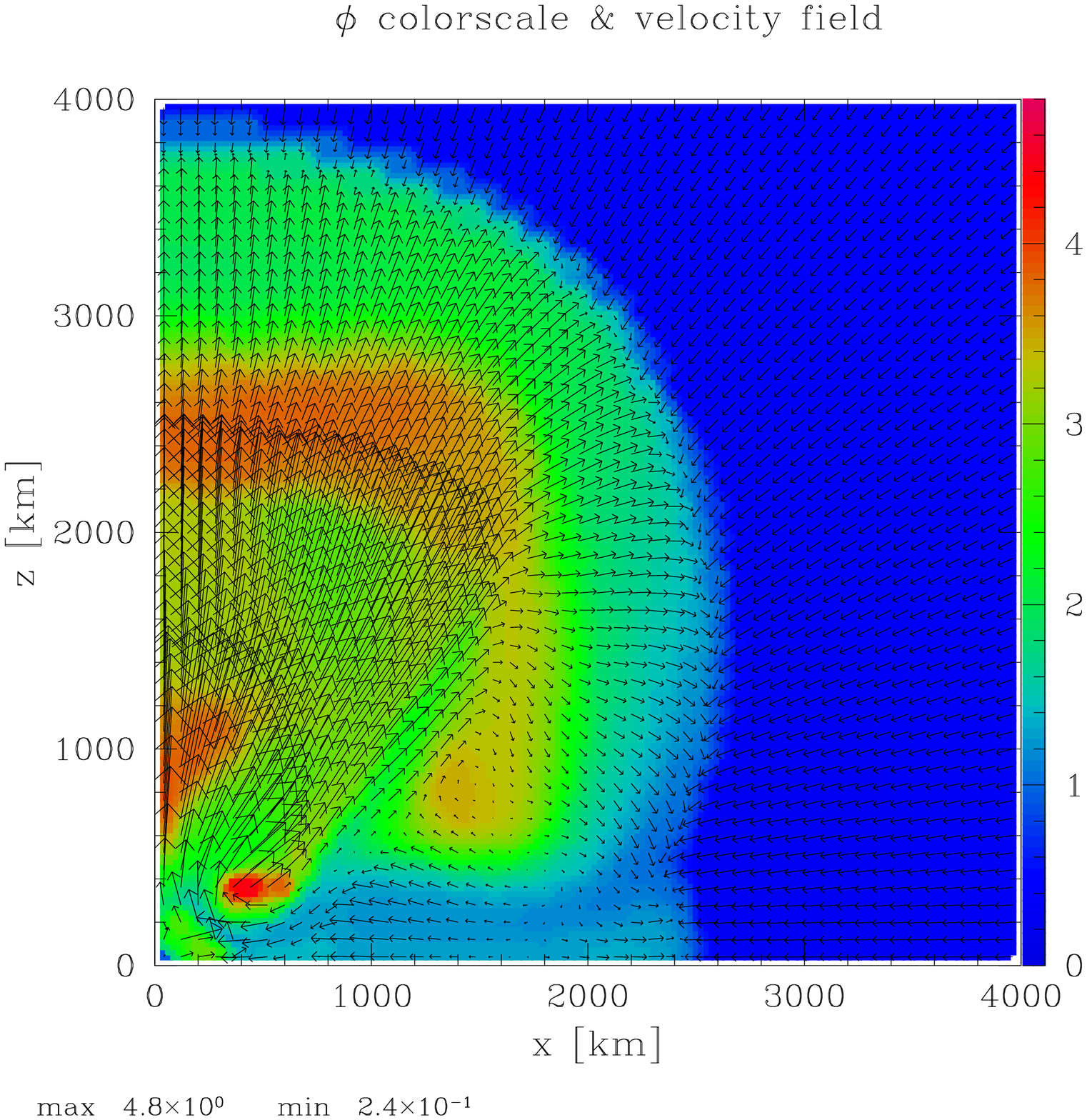}
    \FigureFile(80mm,80mm){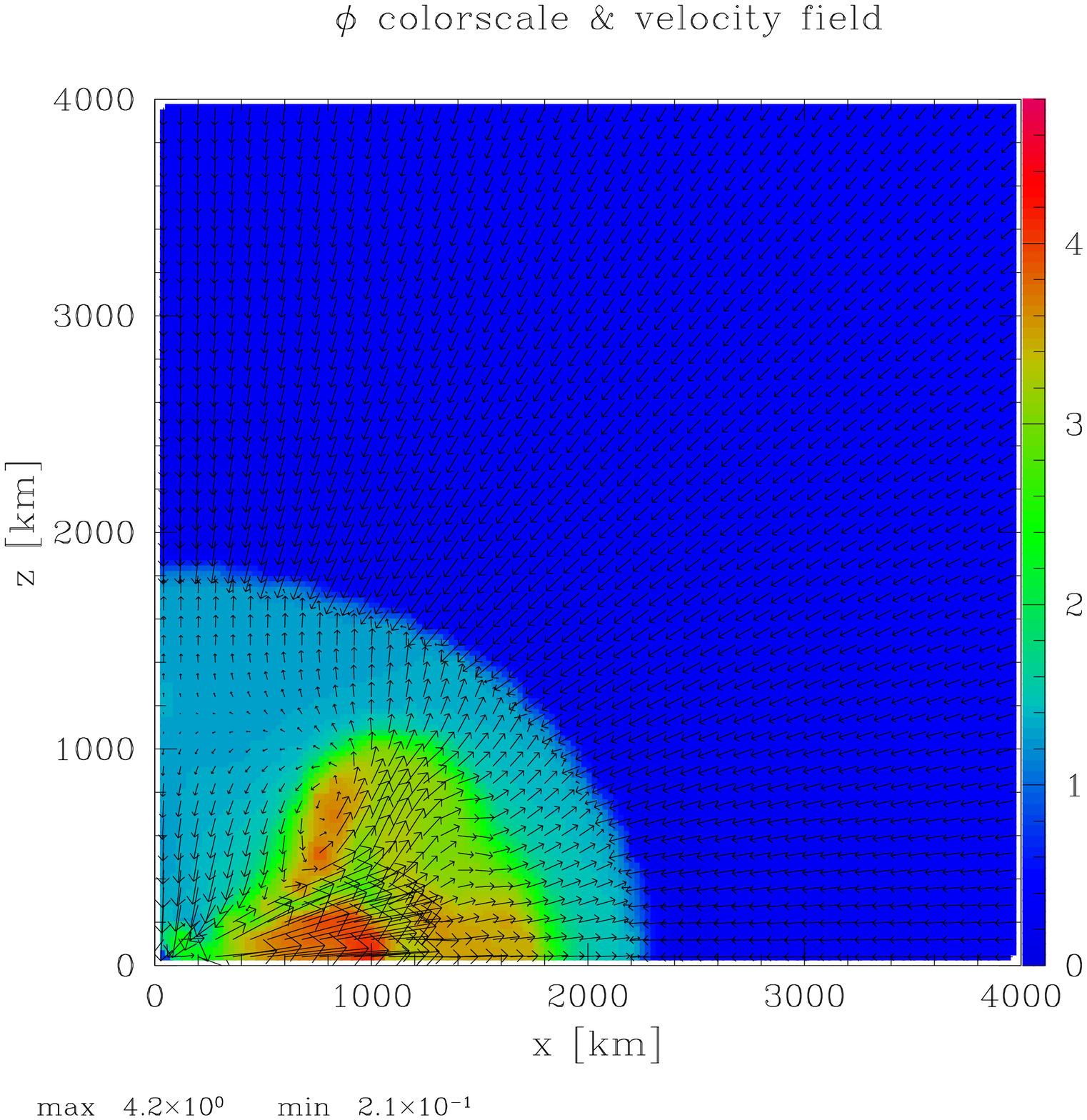}
  \end{center}
  \caption{Color-scale maps of the dimensionless entropy distribution
  and the velocity fields for the two models, pro100 (Left) and obl091
  (Right) at $t \sim 280$ ms after the shock stall.}
  \label{fig:entropy}
\end{figure}

Figure~\ref{fig:entropy} depicts
the color-scale maps of the dimensionless entropy distribution with the
velocity fields for the two models, pro100 and obl091, at $t \sim 280$
ms after the shock stall (for translation to the dimensional entropy,
see Shimizu et al. 2001).  The color boundary between dark blue and
light blue shows the shock front at $r\sim 2500-3800$ km for model
pro100 and at $r\sim 1800-2200$ km for model obl091.  In model pro100,
higher entropy upflows are mainly formed in the region of high
latitudes due to intense neutrino heating on the polar axis.  As the
locally heated matter pushes the shock front, the shock becomes prolate
and largely distorted.  On the other hand, the entropy distribution of
model obl091 shows that the shock front is elongated in a oblate shape
as a result of neutrino heating concentrated on the equatorial plane.
We also notice that the shock front in model pro100 is clearly more
extended than that in model obl091.

\begin{figure}
  \begin{center}
    \FigureFile(80mm,80mm){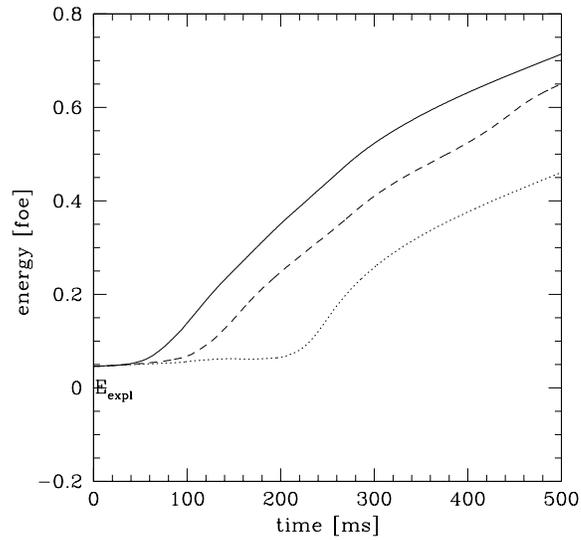}
  \end{center}
  \caption{Evolution of the explosion energy ($E_{\rm expl}$)
  for model pro100 (solid line) and model obl091 (dashed line).  The
  result of the spherical explosion, model sph000 (dotted line), for the same
  $T_{\nu}$ is also shown for comparison.}
  \label{fig:energies}
\end{figure}

The evolution of the explosion energy is illustrated in
Fig.~\ref{fig:energies}.  The result of the spherical model is also
shown for comparison.  We find that the energy gains are substantially
different between the prolate and oblate models.  The increase in the
explosion energy is much larger in model pro100 than that in model
obl091, although both models finally appear to explode.

The difference between the prolate and oblate models is attributed to
the mechanism which was firstly pointed out by \citet{ShEbSaYa01}:
importance of the locally intense neutrino heating.  It should be noted
here that the situations of prolate and oblate explosions are different.
In the former, neutrinos are predominantly emitted along the polar axis.
Because of increased pressure in the locally heated matter near the
pole, the shock wave is partly pushed outwards in the polar direction.
The other part of the shock will follow due to the pressure gradient
along the shock front (Rankine-Hugoniot relation).  In
Fig.~\ref{fig:energies}, one sees that the explosion energy begins to
increase at the position of $t\sim 40$ ms.  It is confirmed that this
point corresponds to the time when the stalled shock starts to move
outwards.

On the other hand, in the case of the oblate explosion, the neutrino
radiation is anisotropic but its peak is not located point-like; the
heated region is {\it disk-like} over the equatorial plane.  The
efficiency of anisotropic neutrino heating of the oblate model is
considerably reduced compared with that of the prolate (local) one.
This is naturally expected from the physical mechanism claimed in
\citet{ShEbSaYa01}.

Note that the neutrino heating rate itself is proportional to
$T_{\nu}^{6}$ and there are no remarkable difference in the neutrino
heating rate between the prolate and oblate models.  The neutrino
cooling rate, however, behaves as $\sim T_{\rm m}^{6}$, where
$T_{\rm m}$ is the matter temperature.  As a result of the earlier
shock revival, $T_{\rm m}$ in model pro100 is quickly reduced.  The
neutrino cooling for the prolate model is therefore largely suppressed
as the decrease of matter temperature in the course of the shock
expansion.  Thus, the explosion energy in model pro100 becomes
larger than that in model obl091 (see Fig.~\ref{fig:energies}).

Up to here, we have shown only the results of two models (pro100 and
obl091) which have the same degree of anisotropy in the neutrino
radiation.  We need not to study many models for the prolate and oblate
explosions with the same degree of anisotropy.  From systematic
investigation in the previous papers (Shimizu et al. 2001; Madokoro et
al. 2003), it is learned that we will obtain the same result if we
examine many models with various degrees of anisotropy for the prolate
and oblate neutrino radiation: the prolate explosion is always more
powerful than the oblate one with the same degree of anisotropy for
a given luminosity.

\begin{figure}
  \begin{center}
    \FigureFile(80mm,80mm){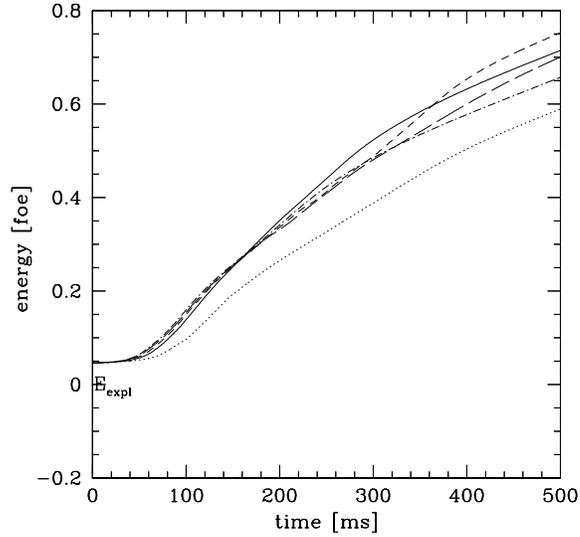}
  \end{center}
  \caption{Evolution of the explosion energy ($E_{\rm expl}$)
  for the prolate models: pro050 (dotted line), pro100 (solid line),
  pro150 (short-dashed line), pro200 (long-dashed line), and
  pro250 (dot-dashed line).  The result of model pro300
  is close to that of model pro250 and hence we omit it.}
  \label{fig:energiespro}
\end{figure}

\begin{figure}
  \begin{center}
    \FigureFile(80mm,80mm){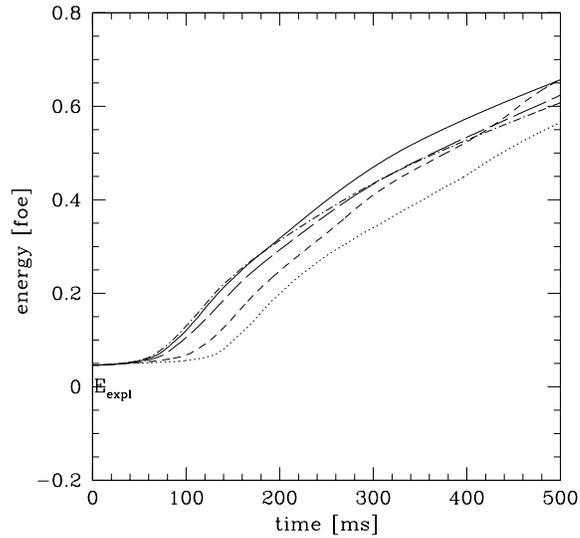}
  \end{center}
  \caption{Evolution of the explosion energy ($E_{\rm expl}$)
  for the oblate models: obl048 (dotted line), obl091 (short-dashed line),
  obl130 (long-dashed line), obl167 (solid line), and
  obl200 (dot-dashed line).  The result of model obl231
  is close to that of model obl200 and hence we omit it.}
  \label{fig:energiesobl}
\end{figure}

\begin{figure}
  \begin{center}
    \FigureFile(80mm,80mm){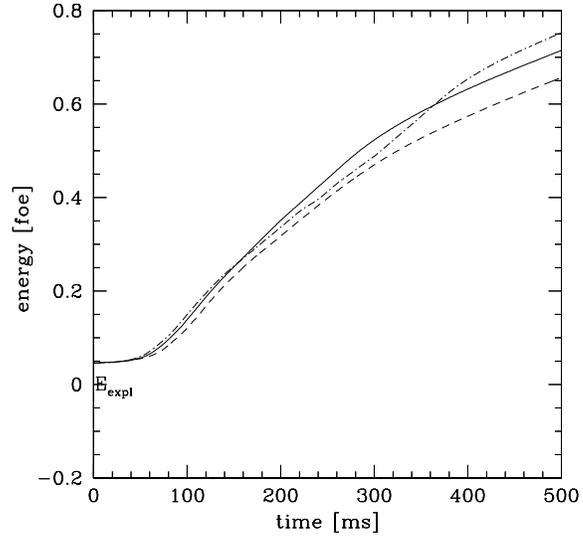}
  \end{center}
  \caption{Evolution of the explosion energy ($E_{\rm expl}$)
  for the saturated models: pro100 (solid line), pro150(dot-dashed line),
  and obl167 (dashed line).}
  \label{fig:energiessat}
\end{figure}

Next, we investigate the cases in which the prolate and oblate
explosions have different degree of anisotropy.  We have found that
the effect of anisotropy becomes more powerful in both the prolate
and oblate models as the degree of anisotropy in the neutrino radiation
increases: the larger the degree of anisotropy is, the larger the
explosion energy becomes.  It was found, however, that this effect
saturates around a certain degree of anisotropy (see discussion in
Shimizu et al. 2001 and Madokoro et al. 2003).  Figures~\ref{fig:energiespro}
and \ref{fig:energiesobl} show the saturation properties of the anisotropic
neutrino radiation for the prolate and oblate models, respectively.
We notice that, at a later stage of evolution, model pro100
($t<360$ ms) or pro150 ($t>360$ ms) shows the largest explosion energy among
the prolate models (Fig.~\ref{fig:energiespro}), while the explosion energy
of model obl167 becomes largest among the oblate models
(Fig.~\ref{fig:energiesobl}).  This means that the saturation of the effect
of anisotropic neutrino radiation comes at $l_{z}/l_{x} \sim 1.1 - 1.15 $
for the prolate explosion and $l_{x}/l_{z} \sim 1.2$ for the oblate explosion.
We also confirmed that the
explosion energy in the prolate model of $l_{z}/l_{x}=1.1$ is larger
than that in the saturated oblate model of $l_{x}/l_{z}=1.2$.  This is
explicitly illustrated in Fig.~\ref{fig:energiessat}.  In other
words, the prolate explosion of $l_{z}/l_{x}=1.1$ is always more
energetic than any other oblate models even if the degree of anisotropy
in the neutrino radiation becomes larger.  We therefore conclude that
the prolate explosion is generally more energetic than the oblate one
for a given neutrino luminosity.

For the purpose of clarifying the sensitivity of the supernova problem
on the neutrino heating, we have also carried out simulations of less
luminous ($T_{\nu}=4.65$ MeV) and more energetic ($T_{\nu}=4.75$ MeV)
models.  When $T_{\nu}=4.65$ MeV, we found that the prolate model
(pro100T465) does explode while the oblate one (obl091T465)
fails.  Note that the difference between
4.70 and 4.65 MeV is only 1\%.  In contrast, the difference between
the prolate and oblate models becomes much smaller when $T_{\nu}$ is
increased to 4.75 MeV (models pro100T475 and obl091T475).
These results indicate that the effect of local neutrino heating becomes
more pronounced and therefore more important when the total neutrino
luminosity is lower.

The spherical model also finally explodes when $T_{\nu} = 4.70$ MeV or
higher (see Fig.~\ref{fig:energies}).  This simply means that any models will
explode irrespective of the degree of anisotropy if the luminosity is
sufficiently high.  However, we should keep in mind that the total neutrino
luminosity can not be simply increased due to the problem of small
mass of the protoneutron star and that of Ni overproduction, especially
in the case of essentially spherical models (Burrows et al. 1995;
Janka et al. 1996).  On the other hand, our model based on anisotropic
neutrino radiation can explode at lower neutrino luminosity, and therefore
could solve the problems described above.

In our previous article (Madokoro et al. 2003), we made a comparison
between globally anisotropic (prolate) explosions and fluctuated
explosions.  We concluded in that paper that globally anisotropic
neutrino radiation is more effective than the fluctuated neutrino
radiation to increase the explosion energy for a given neutrino
luminosity.  In addition, in the present study we have found that the
prolate neutrino heating produces more energetic explosion than the
oblate one.  Combined with the result of \citet{ShEbSaYa01}, we conclude
that the globally anisotropic (prolate) neutrino radiation is the most
effective way of increasing the explosion energy among various types
of explosions investigated in these studies when the total neutrino
luminosity is given.  These results support the statement
made by \citet{ShEbSaYa01}:  globally anisotropic neutrino radiation,
or locally intense neutrino heating is of great importance to produce
a successful explosion.

As we have shown, the locally intense neutrino heating along the polar
axis produces a robust explosion in which the shock front is deformed
in a prolate form.  Our results may be related to the observation that
the expanding materials of SN1987A has a prolate geometry (see Wang
et al. 2002, H\"oflich et al. 2002).

\section{Conclusion}
We have performed 2-D numerical simulations of core-collapse
supernova explosions with prolate and oblate neutrino radiation fields.
We found that these two models give different results in the shock
position at the same evolutionary stage of explosion, the explosion
energy, and the geometry of asymmetric explosion.
It is also found that the prolate explosion yields larger energy gain
than the oblate one for a given neutrino luminosity.
We can expect from the mechanism suggested by Shimizu et al.
(2001) that the prolate explosion is more energetic than the oblate one
when the degree of anisotropy in the neutrino radiation is the same
between these two.  It is not obvious, however, whether the prolate explosion
is always more powerful than the oblate one when the degree of anisotropy
varies independently in these two models.  Our numerical study in this paper
has clarified that the prolate explosion ($l_{z}/l_{x}=1.1$) is always
more powerful than the oblate ones irrespective of the degree of anisotropy.

Moreover, we found that the difference between the prolate and
oblate explosions becomes prominent when the total neutrino luminosity
is low.  This means that the local (prolate) neutrino radiation
becomes more important to increase the explosion energy as the total
neutrino luminosity decreases.  Because we can not simply increase the total
luminosity to explain the observed explosion energy due to the problem of small
mass of the protoneutron star and that of Ni overproduction, especially
in the case of essentially spherical models (Burrows et al. 1995;
Janka et al. 1996).  Therefore, we conclude that the local (prolate) neutrino
radiation is of great importance in actual supernova explosions.

One of the authors showed for the first time that the locally intense
neutrino heating along the axis of symmetry produces more powerful
explosion than the spherical explosion when the total neutrino
luminosity is given \citep{ShEbSaYa01}.  Our results in this article,
together with the results of our previous work (Madokoro et al. 2003)
show that the prolate explosion is the most effective way of increasing
the explosion energy when the total neutrino luminosity is low
among various types of explosions examined in
these studies.  Our results are suggestive of the observation that the
expanding materials of SN1987A has been observed to be deformed in a
prolate form (see Wang et al. 2002, H\"oflich et al. 2002).

\begin{table}
  \caption{Simulated Models.  Note that the value of
  $l_{z}/l_{x}=1+c_{2}$ corresponds to the ratio of the neutrino flux
  along the polar axis and that on the equatorial plane for an observer
  far from the neutrinosphere.  The value of $a$ is calculated from
  equation (\ref{eqn:c2a}).}
  \label{tab:first}
  \begin{center}
    \begin{tabular}{|l|l|l|l|l|}
      \hline
      Model & $l_{z}/l_{x}$ & \hspace*{1em}$c_{2}$ & \hspace*{1em}$a$ & $T_{\nu}$(MeV) \\
      \hline
      \hline
      sph000 & $1.00$ & $0$ & $0$ & 4.70 \\
      \hline
      pro050 & $1.05$ & $+0.050$ & $+0.211$ & 4.70 \\
      pro100 & $1.10$ & $+0.100$ & $+0.444$ & 4.70 \\
      pro150 & $1.15$ & $+0.150$ & $+0.706$ & 4.70 \\
      pro200 & $1.20$ & $+0.200$ & $+1.000$ & 4.70 \\
      pro250 & $1.25$ & $+0.250$ & $+1.333$ & 4.70 \\
      pro300 & $1.30$ & $+0.300$ & $+1.714$ & 4.70 \\
      \hline
      obl048 & 1/1.05 & $-0.048$ & $-0.182$ & 4.70 \\
      obl091 & 1/1.10 & $-0.091$ & $-0.333$ & 4.70 \\
      obl130 & 1/1.15 & $-0.130$ & $-0.462$ & 4.70 \\
      obl167 & 1/1.20 & $-0.167$ & $-0.571$ & 4.70 \\
      obl200 & 1/1.25 & $-0.200$ & $-0.667$ & 4.70 \\
      obl231 & 1/1.30 & $-0.231$ & $-0.750$ & 4.70 \\
      \hline
      sph000T465 & $1.00$ & $0$ & $0$ & 4.65 \\
      sph000T475 & $1.00$ & $0$ & $0$ & 4.75 \\
      pro100T465 & $1.10$ & $+0.100$ & $+0.444$ & 4.65 \\
      pro100T475 & $1.10$ & $+0.100$ & $+0.444$ & 4.75 \\
      obl091T465 & 1/1.10 & $-0.091$ & $-0.333$ & 4.65 \\
      obl091T475 & 1/1.10 & $-0.091$ & $-0.333$ & 4.75 \\
      \hline
    \end{tabular}
  \end{center}
\end{table}



\end{document}